\documentclass[twocolumn,showpacs,superscriptaddress,amsmath,amssymb,prl]{revtex4-1}

\usepackage[dvips]{graphicx}
\usepackage{dcolumn}
\usepackage{bm}

\begin{document}

\title{Stress-induced magnetic domain selection reveals a conical ground state for the multiferroic phase of Mn$_{2}$GeO$_{4}$}

\author{J.\,S.\,White}\email{jonathan.white@psi.ch}
\affiliation{Laboratory for Neutron Scattering and Imaging, Paul Scherrer Institut, CH-5232 Villigen, Switzerland}

\author{T.\,Honda}
\affiliation{Division of Materials Physics, Graduate School of Engineering Science, Osaka University, Toyonaka, Osaka 560-8531, Japan}
\affiliation{Condensed Matter Research Center, Institute of Materials Structure Science, High Energy Accelerator Research Organization, Tsukuba 305-0801, Japan}

\author{R.\,Sibille}
\author{N.\,Gauthier}
\affiliation{Laboratory for Scientific Developments and Novel Materials, Paul Scherrer Institut, CH-5232 Villigen, Switzerland}
\author{V.\,Dmitriev}
\affiliation{SNBL at ESRF, Polygone Scientifique Louis N\'{e}el, 6 rue Jules Horowitz, 38000 Grenoble, France}

\author{Th.\,Str\"{a}ssle}
\author{Ch.\,Niedermayer}
\affiliation{Laboratory for Neutron Scattering and Imaging, Paul Scherrer Institut, CH-5232 Villigen, Switzerland}

\author{T.\,Kimura}
\affiliation{Division of Materials Physics, Graduate School of Engineering Science, Osaka University, Toyonaka, Osaka 560-8531, Japan}

\author{M.\,Kenzelmann}
\affiliation{Laboratory for Scientific Developments and Novel Materials, Paul Scherrer Institut, CH-5232 Villigen, Switzerland}

\date{\today}
\begin{abstract}
At ambient pressure $P$ and below 5.5~K, olivine-type Mn$_{2}$GeO$_{4}$ hosts a multiferroic (MF) phase where a multi-component, i.e. multi-$k$ magnetic order generates spontaneous ferromagnetism and ferroelectricity (FE) along the \textbf{c}-axis. Under high $P$ the FE disappears above 6~GPa, yet the $P$ evolution of the magnetic structure remained unclear based on available data. Here we report high-$P$ single crystal neutron diffraction experiments in the MF phase at $T=$~4.5~K. We observe clearly that the incommensurate spiral component of the magnetic order responsible for FE varies little with $P$ up to 5.1~GPa. With support from high $P$ synchrotron x-ray diffraction measurements at room temperature ($T$), the $P$ driven suppression of FE is proposed to occur as a consequence of a crystal structure transition away from the olivine structure. In addition, in the low $T$ neutron scattering experiments an emergent non-hydrostatic $P$ component, i.e. a uniaxial stress, leads to the selection of certain multi-$k$ domains. We use this observation to deduce a double-$k$ conical magnetic structure for the ambient $P$ groundstate, this being a key ingredient for a model description of the MF phase.
\end{abstract}


\maketitle
\section{INTRODUCTION}
\label{sec:1Int}
Multiferroic (MF) materials are exciting systems in which to study the basic interplay between structural, magnetic and electric degrees of freedom. For clean, \emph{in situ} tuning between phases with contrasting MF properties, high pressure ($P$) is a key experimental parameter~\citep{Gil14}. For the magnetically-driven MFs where symmetry-breaking magnetism generates ferroelectricity (FE) directly~\citep{Tok14}, high $P$ studies have contributed novel results in recent years. Examples include the observed $P$ driven reversal of the direction of ferroelectric polarization in YMn$_{2}$O$_{5}$~\citep{Cha08,Koz15}, the theoretical expectation for a room temperature ($T$) MF state in CuO under high $P$~\citep{Roc13}, and the observation of a $P$ driven magnetoelectric (ME) phase transition in TbMnO$_{3}$ into a high $P$ state hosting a huge spin-driven electric polarization~\citep{Aoy14}. These studies show the potential for high $P$ to give access to novel physics involving multiferroicity.

\begin{figure}
\includegraphics[width=0.48\textwidth]{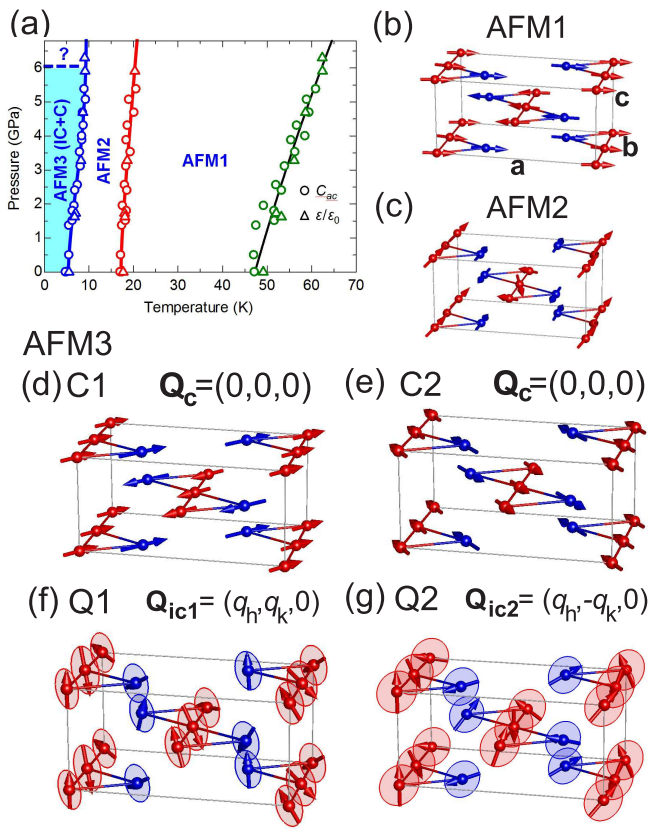}
\caption{(Color online) (a) $P$-$T$ phase diagram for Mn$_{2}$GeO$_{4}$ determined by various bulk measurement techniques as reported in Ref.~\onlinecite{Hon14}. Panels (b)-(g) show the magnetic structures in the various phases as determined at ambient $P$~\citep{Whi12}. The magnetic Mn$^{2+}$ ($S$=5/2) ions, occupy two different sites~\citep{Hag00}; the 4($a$) site shown in red, and the 4($c$) site shown in blue. Arrows represent magnetic moment directions. Panels (b) and (c) respectively show the magnetic structures for the AFM1 and AFM2 phases. Panels (d) and (e) show two possible C domains that exist in the MF AFM3 phase. Panels (f) and (g) show possible spiral structures that may exist within each of the two possible IC $k$-domains in the MF phase. In each of the last two panels, envelopes around the moments denote a common spin rotation plane.}
\label{fig:Mag_Structs}
\end{figure}

Neutron scattering is a powerful probe of the microscopic magnetic correlations in spin-driven MF materials, though relatively few high $P$ neutron studies are reported to date~\citep{Gil14,Kim08,Koz10,Ter14,Deu15,Ter16}. Here we use the technique to explore the high $P$ evolution of the magnetism in the MF state of the orthorhombic ($Pnma$) olivine-type Mn$_{2}$GeO$_{4}$ (MGO)~\citep{Cre70}. As shown in Fig.~\ref{fig:Mag_Structs}(a), at ambient $P$ MGO displays three magnetic phases as a function of $T$: $T_{\rm N1}$ = 47~K $>$ AFM1 $>$ $T_{\rm N2}$ = 17~K $>$ AFM2 $>$ $T_{\rm N3}$ = 5.5~K $>$ AFM3~\citep{Whi12,Hon12,Vol13}. The AFM1 and AFM2 phases are paraelectric and host simple commensurate (C) antiferromagnetic structures described by the propagation vector $\textrm{Q}_{\rm c}=(0~0~0)$ [Figs.~\ref{fig:Mag_Structs}(b) and (c)].

The AFM3 phase hosts a spin-driven MF state with spontaneous ferromagnetism (FM) and ferroelectric polarization both along the \textbf{c}-axis~\citep{Whi12,Hon12}. Using neutron diffraction it was shown that this MF state hosts both C and incommensurate (IC) magnetic orders simultaneously~\citep{Whi12}. The C order has a propagation vector $\textrm{Q}_{\rm c}=(0~0~0)$ and is described by a combination of two irreducible representations $\Gamma^{1}+\Gamma^{3}$. This is consistent with a magnetic point group symmetry $2/m$ and a monoclinic axis along the \textbf{c}-direction that allows for FM. Figs.~\ref{fig:Mag_Structs}(d) and (e) respectively show two possible distinct C domains C1 ($\Gamma^{1}+\Gamma^{3}$) and C2 ($\Gamma^{1}-\Gamma^{3}$).

The IC order is a doubly-IC spin spiral with a general propagation vector $\textrm{Q}_{\rm ic}=(q_{h}~q_{k}~0)$, where $q_{h}=$~0.136 and $q_{k}=$~0.211 at ambient $P$~\citep{Whi12}. This spiral order generates the FE, and is describable by a sum of two corepresentations of $\textrm{Q}_{\rm ic}$, $D^{1}+D^{2}$~\citep{Whi12}. Possible spiral structures in each of the two IC $k$-domains, $\textrm{Q}_{\rm ic1}=(q_{h}~q_{k}~0)$ (Q1) and $\textrm{Q}_{\rm ic2}=(q_{h}~-q_{k}~0)$ (Q2) are shown in Figs.~\ref{fig:Mag_Structs}(f) and (g), respectively. The IC modulation lowers the magnetic point group symmetry further to $2$; a polar point group with just a two-fold rotation about the \textbf{c}-axis that allows the FM and FE to co-exist along this direction~\citep{Whi12}.

The coherent superposition of the C and IC magnetic modulations to form multi-$k$ structures is evidenced by both bulk measurements and reciprocal space neutron scattering~\citep{Whi12,Hon12}. In addition, recent second-harmonic generation measurements prove the real-space coexistence of ferromagnetic and ferroelectric domains~\citep{Leo16}. The formation of multi-$k$ domains thus provides a microscopic basis for a coupling between the bulk properties of FM and FE. It is proposed that such a coupling may be mediated by Dzyaloshinskii-Moriya interactions that are unique to the MF phase~\citep{Whi12}.

The motivation for the present high $P$ study of MGO is the following. In the low temperature (LT) MF phase of MGO, high $P$ bulk measurements show the FE along the \textbf{c}-axis to disappear for $P>P^{\ast}_{\rm LT}\approx6$~GPa~\citep{Hon14}. To explain this observation, the $P$ evolution of the magnetism was studied by high $P$ powder neutron diffraction up to 5.3~GPa~\citep{Hon14}. The data clearly showed the $\Gamma^{1}+\Gamma^{3}$ symmetry of the C order to survive up to 5.3~GPa. However, due to both peak overlap and the weak scattering from the IC peaks, the $P$ evolution of the IC order was unclear. In particular it could not be determined if the IC order became suppressed already at a $P$ lower than $P^{\ast}_{\rm LT}$~\citep{Hon14}. From a general viewpoint, it is of interest to establish the nature of the presumed suppression of IC order as $P\rightarrow P^{\ast}_{\rm LT}$. A continuous suppression could hint at MGO being an interesting system for studying the critical properties of the MF transition.

Here we report high $P$ scattering measurements of the structure and magnetism in MGO. Using high $P$ neutron diffraction to study single crystal samples at low $T$, we avoid the problems that hampered the interpretation of the previous powder diffraction experiments~\citep{Hon14}. Consequently the $P$ dependence of the magnetic order is easily determined up to 5.1~GPa, the highest $P$ achieved. Despite not quite reaching $P^{\ast}_{\rm LT}$, by combining the results with room $T$, high $P$ synchrotron x-ray diffraction (SXRD) measurements up to 10~GPa, we propose a consistent picture for the low $T$ transition at $P^{\ast}_{\rm LT}$. In addition, the low $T$ neutron experiments evidence an emergent anisotropic stress $P$ component at higher $P$s that leads to the stabilization of particular multi-$k$ domains in the MF AFM3 phase. We use this observation to propose the full multi-$k$ magnetic structures of the MF domains at ambient $P$, this being a key feature for a model description of the MF ground state.


\section{EXPERIMENTAL METHOD}
\label{sec:2Int}
High $P$ SXRD experiments were carried out at the Swiss-Norwegian Beamline (SNBL) at the ESRF, Grenoble, France. The experiment was done on a powder sample of MGO that was obtained from ground single crystals prepared by the floating zone method. The sample was loaded into a diamond anvil cell (DAC) with a pressure transmitting medium (PTM) of ethanol-methanol. Some ruby chips were also added; this allowed the \emph{in situ} measurement of the sample $P$ using the ruby fluorescence method. X-ray powder diffraction datasets were collected using a monochromatic beam of wavelength $\lambda=$~0.69563~\AA, and a 2D detector (Pilatus 2M, Dectris). 2D images showing good powder averaging were integrated and then converted into 1D diffraction patterns of intensity vs. diffraction angle. The resulting datasets cover the range $0.5\leq Q \leq6.4$~\AA$^{-1}$ with a resolution $\delta Q\approx0.01$~\AA$^{-1}$. The diffraction patterns were analyzed using the FullProf suite and the peaks were modeled by pseudo-Voigt functions (peakshape function 5)~\citep{Rod93}.

Low $T$ and high $P$ single crystal neutron diffraction experiments made use of opposed-anvil techniques and a Paris-Edinburgh (PE) VX5 press~\citep{Klo05}. The press had a dedicated cryocooler which provided a sample base $T$ of 4.5~K. A MGO single crystal sample of approximate dimension 2 x 2 x 2 mm$^{3}$ and mass 56~mg was cut and aligned with [100] and [010] in the horizontal scattering plane. The crystal was enveloped entirely within a pre-pressed polycrystalline Pb matrix that was itself positioned inside a bespoke CuBe gasket. The profile of the matrix and gasket assembly was shaped carefully so that it matched the sample space available between the two anvils. By embedding the sample within a soft Pb matrix, i) the sample alignment could be maintained under pressurization, ii) the soft Pb provides a quasi-hydrostatic PTM, and iii) the sample $P$ could be determined in-situ by tracking the $P$  and $T$ dependent lattice constant of the Pb, and using the recently determined equation of state~\citep{Str14}.

\begin{figure}
\includegraphics[width=0.48\textwidth]{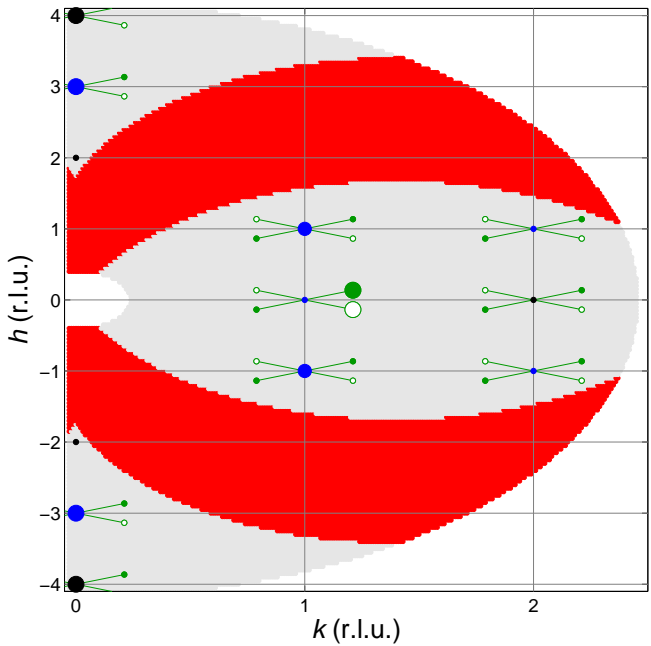}
\caption{(Color online) Sketch of reciprocal space in the ($h$,$k$,0) horizontal plane for both the MF phase of MGO and the experimental setup described in the text. The red regions correspond to inaccessible portions of reciprocal space due the geometrical constraints imposed by the pillars of the PE press. The light gray region defines the overall region of accessible reciprocal space. Filled black symbols, and filled blue symbols respectively denote positions where scattering from the nuclear structure, or C magnetism is observed. Empty and filled green symbols respectively denote positions where magnetism due to IC domains Q1 and Q2 are observable. For each origin of scattering, a larger symbol size denotes where stronger scattering is observed.}
\label{fig:Exp}
\end{figure}

The PE Press was installed at the RITA-II instrument located at the Swiss spallation neutron source, SINQ, PSI, Switzerland. Elastic neutron diffraction measurements were carried out using an incident neutron energy of 4.6~meV. A cold Be filter was placed between the sample and analyser to suppress the second-order contamination of the neutron beam. Diffraction measurements were performed mostly at the base $T$ of 4.5~K, and the $P$ changes carried out at elevated $T$ above 180~K.

The PE press construction has two openings of 140$^{\circ}$ that are separated on each side by two 40$^{\circ}$ pillars that can block the incoming or outgoing neutron beam~\citep{Klo05}. The pillars thus impose a restriction on the accessible range of reciprocal space, but this can be negated by suitably orienting the sample-gasket ensemble within the PE press. Figure~\ref{fig:Exp} shows the accessible reciprocal space for the chosen sample orientation, and which structural and magnetic peaks could be accessed in our experiments.

\section{RESULTS AND ANALYSIS}
\label{sec:3Res}
\subsection{Crystal structure}
\label{sec:3_1_Crys}
\begin{figure}
\includegraphics[width=0.48\textwidth]{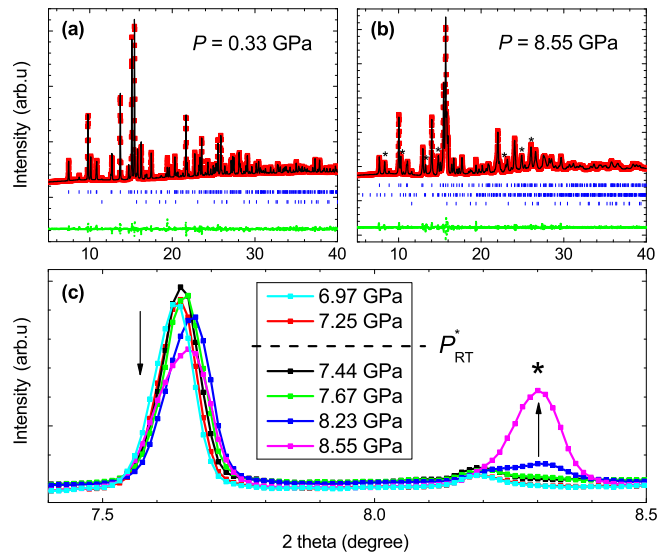}
\caption{(Color online) SXRD patterns from MGO measured at room $T$ and (a) $P=$~0.33~GPa and (b) $P=$~8.55~GPa. In both panels, the observed (red), calculated (black), and difference (green) profiles are shown. The top row of blue ticks show the expected Bragg peaks due to the olivine ($\alpha$-) MGO structure, while the bottom row of ticks denotes the peaks from the ruby. In panel (b), the middle row of blue ticks denotes the expected Bragg peaks due to the high $P$, monoclinic form of the $\beta$-MGO phase. Stars in panel (b) denote peaks from this high $P$ phase. Panel (a) shows a Rietveld refinement of the data with $R_{\rm f}=$~5.58, while panel (b) shows a Lebail refinement with $R_{\rm p}=$~9.28. Panel (c) provides a close look at the diffraction pattern for low scattering angles, and for $P$s around the transition $P$ of $P^{\ast}_{\rm RT}=$~7.35(10)~GPa.}
\label{fig:Lattice_xray}
\end{figure}

We start by reporting the results of the room $T$ SXRD experiments with sample $P$s approaching 10~GPa. Figure~\ref{fig:Lattice_xray}(a) shows the SXRD pattern for the low $P=$~0.33~GPa along with a Rietveld structure refinement including two structural phases; MGO in its ambient $P$ olivine ($Pnma$) form, the so-called $\alpha$-MGO phase~\citep{Mor72}, and the ruby chips used for the \emph{in situ} $P$ determination. As shown in Fig.~\ref{fig:Lattice_xray}(b), at the higher $P=$~8.55~GPa, further Bragg peaks (denoted by stars) are observed in the diffraction pattern. This evidences the emergence of an additional structural phase under high $P$. Figure~\ref{fig:Lattice_xray}(c) shows a closer look at the low scattering angle region for patterns obtained at $P$s close to where the further phase emerges. This phase is first refinable in the data obtained at $P=$~7.44~GPa, and it co-exists with the $\alpha$-MGO phase up to the highest $P=$~9.68~GPa. The co-existence indicates the onset at room temperature (RT) of a first-order structural transition at $P^{\ast}_{\rm RT}=$~7.35(10)~GPa. These observations are broadly consistent with previous high $P$, room $T$ Raman measurements, which reported a co-existence of different structural phases to onset for $P>$~6.7~GPa~\citep{Rey97}.


The SXRD data obtained for $P>P^{\ast}_{\rm RT}$ could not be refined reliably using the Rietveld refinement method. Therefore we used the Lebail refinement technique to identify the space group symmetry and crystal parameters of the high $P$ phase. The new high $P$ phase that coexists with the $\alpha$-MGO phase is best described by a monoclinic unit cell with symmetry $I2/m$, and crystal parameters at 8.55~GPa of $a=$~6.0270(3)~\AA, $b=$~12.1668(4)~\AA, $c=$~8.7232(3)~\AA, and $\beta=$~93.847(2)$^{\circ}$. This unit cell can itself be derived from the orthorhombic one that describes so-called $\beta$-MGO with space group $Imma$ and lattice parameters $a=$~6.025~\AA, $b=$~12.095~\AA, and $c=$~8.752~\AA~\citep{Mor72}.


\begin{figure}
\includegraphics[width=0.48\textwidth]{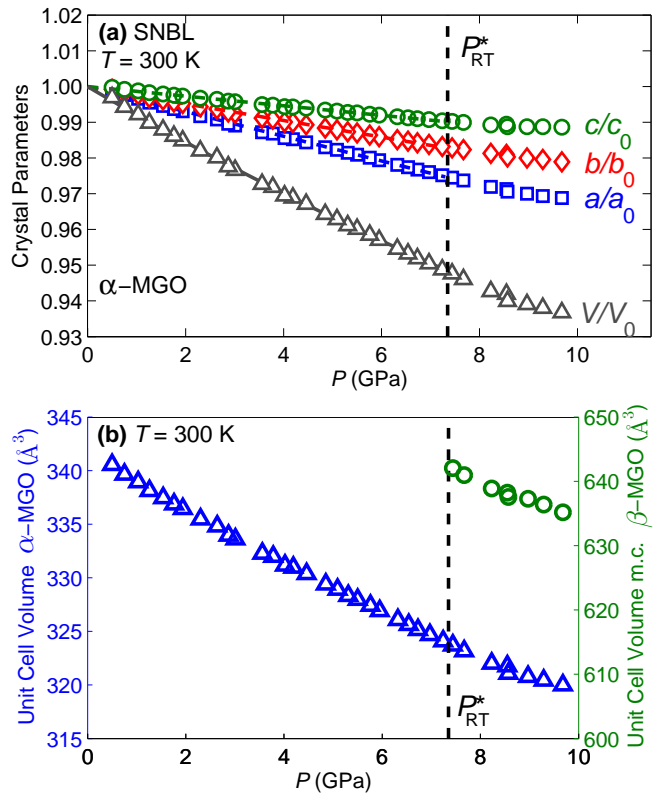}
\caption{(Color online) (a) The $P$ dependence of the crystal structure parameters at room $T$ for the olivine ($Pnma$) $\alpha$-MGO structure that exists at ambient $P$. All data are normalized to values obtained from powder diffraction at ambient $P$; $a_{\rm 0}=$~10.7154(2)~\AA, $b_{\rm 0}=$~6.2951(1)~\AA, $c_{\rm 0}=$~5.0619(1)~\AA, and $V_{\rm 0}=$~341.488(3)~\AA$^{3}$. (b) The $P$ dependence of the unit cell volume for both the olivine $\alpha$-MGO phase (blue triangle) and the monoclinic (m.c.) form of the $\beta$-MGO phase (green circles) - see text for details. In both panels, the dashed black line indicates the $P$ above which the $\beta$-MGO phase is detected in the data.}
\label{fig:Lattice_xray2}
\end{figure}

In Fig.~\ref{fig:Lattice_xray2}(a) we present the $P$ dependence of the normalized crystal lattice parameters of the $\alpha$-MGO phase over the entire $P$ range. From linear fits of the relative changes in the lattice constant for $P$s up to $P^{\ast}_{\rm RT}$, the linear compressibilities $k_{i}=-(1/(a_{i})_{P=0})(da_{i}/dP)_{T}$ are determined to be $k_{a}=0.00344(1)$~GPa$^{-1}$, $k_{b}=0.00232(2)$~GPa$^{-1}$, and $k_{c}=0.00134(1)$~GPa$^{-1}$. The $P$ dependence of the relative change of the unit cell volume $V/V_{0}$, where $V_{0}$ is the unit cell volume at ambient $P$, is well-described by the third-order Birch-Murnaghan equation of state~\citep{Mur37,Bir47,Bir86}. At 300~K, we extract a bulk modulus $B_{0}$=$-V(dP/dV)_{T}$=~123(1)~GPa when using the fixed pressure-derivative of $B_{0}$, $B'$=$(dB_{0}/dP)_{T}$=~4.4 reported for Mg$_{2}$SiO$_{4}$~\citep{Li96}. In Fig.~\ref{fig:Lattice_xray2}(b) we show the $P$ dependence of the unit cell volumes for the two different MGO phases on an absolute scale, both to emphasise their difference, and also the $P$ range of their coexistence.

\begin{figure}
\includegraphics[width=0.48\textwidth]{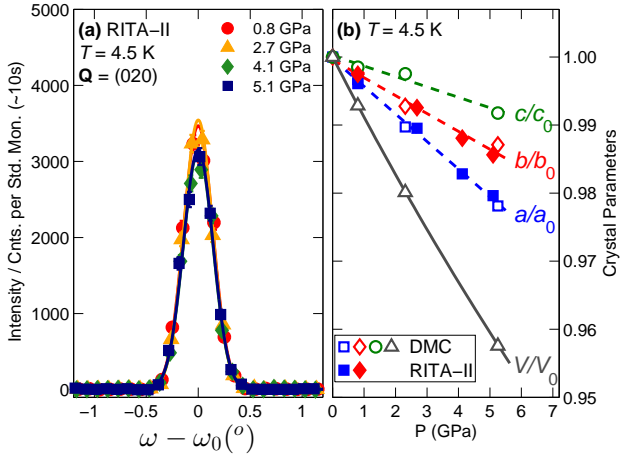}
\caption{(Color online) (a) Scans of the sample rotation angle, $\omega$ through the (020) nuclear peak, rotating the sample relative to the incident neutron beam. The angular dependence is plotted relative to the absolute fitted peak position $\omega_{0}$ for easier comparison between scans done at different $P$s. (b) The $P$ dependence of the crystal structure parameters in the MF phase. Empty symbols denote data extracted from powder neutron diffraction experiments done using the DMC diffractometer at SINQ, PSI~\citep{Hon14}. Filled symbols denote data obtained from single crystal data using RITA-II. All data are normalised to values obtained from powder diffraction at $T=$~4.5~K and ambient $P$; $a_{\rm 0}=$~10.694(2)~\AA, $b_{\rm 0}=$~6.286(1)~\AA, $c_{\rm 0}=$~5.056(1)~\AA, and $V_{\rm 0}=$~339.94(7)~\AA$^{3}$. Dashed lines are linear fits of the $P$ dependent lattice constants. The solid line is an interpolation of the normalised unit cell volume using the third-order Birch-Murnaghan equation of state.}
\label{fig:Lattice}
\end{figure}

Next we turn to aspects of the low $T$ crystal structure obtained from the high $P$ neutron scattering experiments. Figure~\ref{fig:Lattice}(a) shows typical scans obtained from the single crystal sample as the sample angle $\omega$ is rotated through the (020) position. At this position only nuclear scattering is observed. Both the peak width and intensity remain essentially unchanged as $P$ is increased. This respectively indicates there to be no drastic $P$ induced changes in either the crystal mosaicity or apparent lattice symmetry over the studied $P$ range at this $T$. Therefore the lattice distortion induced by the MF transition remains so small that no deviation from the paramagnetic $Pnma$ symmetry is detected.


The $P$ dependence of the low $T$ crystal structure parameters are shown in Fig.~\ref{fig:Lattice}(b). Here data are included from two experiments. Firstly, we include parameters newly extracted from Lebail refinements of the powder neutron diffraction data reported in Ref.~\onlinecite{Hon14}, obtained using the DMC instrument at SINQ, PSI. Secondly, parameters obtained from the single crystal experiments on RITA-II are included. The $a$ and $b$ lattice constants could be determined using data from both experiments, while the $c$ lattice constant and hence unit cell volume $V$, could only be determined from the DMC data. Good agreement is observed between the $P$ dependences of the $a$ and $b$ lattice constants determined from both experiments.

Similarly as for the room $T$ data at high $P$, the low $T$ data shown in Fig.~\ref{fig:Lattice}(b) also evidence an anisotropic lattice compression as $P$ is increased in the MF phase. In this case we find the linear compressibilities for each lattice constant to be $k_{a}=0.0040(2)$~GPa$^{-1}$, $k_{b}=0.0026(2)$~GPa$^{-1}$, and $k_{c}=0.0015(2)$~GPa$^{-1}$, all of these being slightly larger than their corresponding values at room $T$. Concomitantly, from the fit of the $P$ dependence of the normalised unit cell volume $V/V_{0}$ using the third-order Birch-Murnaghan equation of state, we extract a relatively smaller bulk modulus of $B_{0}=$~111(1)~GPa at 4.5~K when using a fixed $B'=$~4.4.

\subsection{Commensurate Magnetic Order}
\label{sec:3_2_CMag}

\begin{figure}
\includegraphics[width=0.48\textwidth]{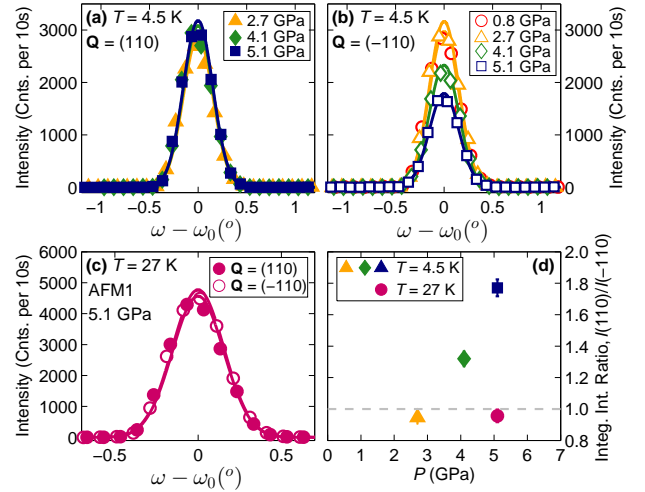}
\caption{(Color online) Scans of the relative sample rotation angle, $\omega$-$\omega_{0}$, through the (a) (110) and (b) ($-$110) magnetic peaks at $T=$~4.5~K and different $P$s. The curves fitted by gaussian lineshapes. (c) Sample angle rotation scans across the (110) and ($-$110) peaks in the AFM1 phase at $T=$~27~K and $P=$~5.1~GPa. Note that more scattering appears at the (110) and ($-$110) positions in the AFM1 phase due to the intrinsic difference between the magnetic structures in the AFM1 and AFM3 phases~\citep{Whi12}. (d) The $P$ dependence of the ratio of integrated intensities for the (110) and ($-$110) peaks, $I(110)/I(-110)$ at different $T$s and $P$s. Integrated intensities are obtained by integrating the area under the lineshapes used to fit the data shown in panels (a)-(c). In all panels, error bars not visible are smaller than the data symbol size.}
\label{fig:C_Mag}
\end{figure}

Next we turn to high $P$ neutron diffraction measurements of the C magnetic order at 4.5~K in the MF phase. Figs.~\ref{fig:C_Mag}(a) and (b) respectively, show the $P$ dependence of the (110) and ($-$110) C magnetic peaks, these being positions where no nuclear scattering is expected. Consistent with the previous high $P$ powder diffraction study~\citep{Hon14}, these data show that the C magnetic order survives up to the highest explored $P$ of 5.1~GPa. We also see that the intensity of the (110) peak varies only weakly with $P$, while the intensity of the ($-$110) peak becomes clearly suppressed for $P>$~2.7~GPa. Since the magnetic symmetry of the C order remains unchanged under high $P$, the magnetic scattering at every $\{$110$\}$ position should be equivalent for equal populations of all C domains. Therefore, inequivalence of the (110) and ($-$110) peak intensities for $P>$~2.7~GPa indicates a $P$ induced change in the populations of the C domains.


The clear difference in the relative intensities of the (110) and ($-$110) peaks at high $P>$~2.7~GPa was found to occur only after cooling into the MF phase. For comparison, Fig.~\ref{fig:C_Mag}(c) shows measurements of both the (110) and ($-$110) C magnetic peaks at 5.1~GPa and 27~K in the paraelectric AFM1 phase. Their integrated intensity ratio, $I(110)/I($-$110)$ shows the two peaks to have equivalent intensities within 5~\%, as expected for the simpler magnetic symmetry of this phase. With Fig.~\ref{fig:C_Mag}(d), we emphasize the different behavior of the ratio $I(110)/I($-$110)$ between the MF and AFM1 phases.


\subsection{Incommensurate Magnetic Order}
\label{sec:3_3_ICMag}
\begin{figure}
\includegraphics[width=0.48\textwidth]{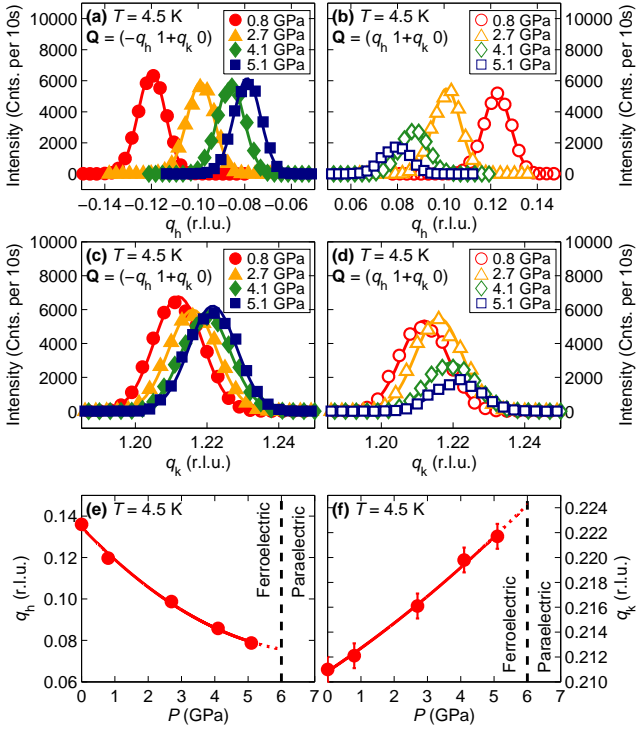}
\caption{(Color online) The $P$ dependence at $T=$~4.5~K of the IC spin spiral component of the magnetic order in the MF phase. Panels (a) and (b) respectively show reciprocal space scans along the $q_{\rm h}$ direction of the peaks \textbf{Q}=($-q_{\rm h}$ 1+$q_{\rm k}$ 0) and \textbf{Q}=($-q_{\rm h}$ 1+$q_{\rm k}$ 0). Panels (c) and (d) show scans through the same peaks but instead along the $q_{\rm k}$ direction. All curves are fitted by gaussian lineshapes, and the $P$ dependence of the mean fitted peak positions in $q_{\rm h}$ and $q_{\rm k}$ are plotted in panels (e) and (f). In the latter two panels, solid lines are guides for the eye and dashed lines denote the ferroelectric to paraelectric boundary at $P^{\ast}_{\rm LT}\approx 6$~GPa determined from bulk measurements~\citep{Hon14}. In all panels, error bars not visible are smaller than the size of the data symbol.}
\label{fig:IC_Mag}
\end{figure}

Figures~\ref{fig:IC_Mag}(a)-(f) summarize $P$ dependent neutron diffraction measurements of the IC order in the MF phase. Figures~\ref{fig:IC_Mag}(a) and \ref{fig:IC_Mag}(c), respectively, show scans along the $h$ and $k$ directions in reciprocal space through the IC magnetic peak \textbf{Q}=($-q_{\rm h}$ 1+$q_{\rm k}$ 0). Here \textbf{Q}=\textbf{G}+\textbf{Q}$_{\rm ic2}$, where \textbf{G} is the reciprocal lattice vector (010), and \textbf{Q}$_{\rm ic2}$ is the propagation vector due to IC $k$-domain Q2. From these data, we find that the \emph{magnitude} of the $q_{\rm h}$ component of the incommensuration decreases with increasing $P$, while that of the $q_{\rm k}$ component increases slightly. Similar measurements across the IC magnetic peak \textbf{Q}=($q_{\rm h}$ $1+q_{\rm k}$ 0) in the Q1 domain are shown in Figs.~\ref{fig:IC_Mag}(b) and \ref{fig:IC_Mag}(d). We observe that the $P$ dependences of $q_{\rm h}$ and $q_{\rm k}$ for this peak are consistent with those expected for when the Q1 and Q2 domains remain configurational $k$-domains across the entire $P$ range. In Figs.~\ref{fig:IC_Mag}(e) and \ref{fig:IC_Mag}(f) we plot the overall $P$ dependence of the $q_{\rm h}$ and $q_{\rm k}$ components of the IC order. Each component varies monotonically as $P$ increases, though neither becomes close to an obvious C value as $P\rightarrow P^{\ast}_{\rm LT}$. Indeed, the extrapolation of the data suggests that the IC order survives easily until $P^{\ast}_{\rm LT}$$\approx6$~GPa.

\begin{figure}
\includegraphics[width=0.48\textwidth]{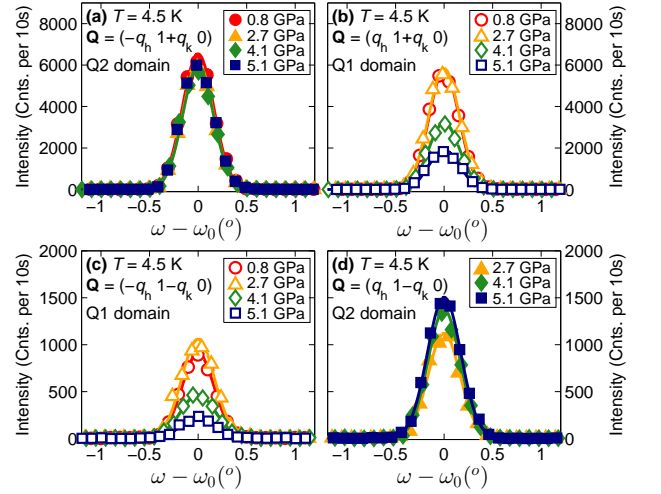}
\caption{(Color online) Scans of the relative sample rotation angle, $\omega$-$\omega_{0}$, through the IC magnetic peaks (a) ($-q_{\rm h}$ $1+q_{\rm k}$ 0), (b) ($q_{\rm h}$ $1+q_{\rm k}$ 0), (c) ($-q_{\rm h}$ $1-q_{\rm k}$ 0) and (d) ($q_{\rm h}$ $1-q_{\rm k}$ 0) at $T=$~4.5~K and different pressures. The IC peaks shown in panels (a) and (d) ((b) and (c)) belong the Q2 (Q1) domain. All curves are fitted by gaussian lineshapes and error bars not visible are smaller than the size of the data symbol.}
\label{fig:IC_Mag_a3}
\end{figure}

Figures~\ref{fig:IC_Mag}(a)-\ref{fig:IC_Mag}(d) also show that the relative intensities of the different IC peaks vary when $P>$~2.7~GPa. This corresponds to the same $P$ range for which the ($-$110) C peak becomes suppressed relative to the (110) C peak, and here it similarly indicates a $P$ driven change in the relative populations of the two IC $k$-domains. To quantify this effect more accurately, rotation angle ($\omega$) scans were done to obtain measures of the peak integrated intensities. Figs.~\ref{fig:IC_Mag_a3}(a) and \ref{fig:IC_Mag_a3}(b) respectively show the $\omega$ scans for the ($-q_{\rm h}$ $1+q_{\rm k}$ 0) peak in the Q2 IC domain, and the ($q_{\rm h}$ $1+q_{\rm k}$ 0) peak in the Q1 domain. In zero fields and at ambient $P$, these two peaks have equivalent intensities when the Q1 and Q2 domains populate the sample equally~\citep{Whi12}. Therefore, the data indicate that for $P>$~2.7~GPa the Q2 domain is more populated than the Q1 domain.

This picture is supported by further measurements done at different IC positions, as shown in Figs.~\ref{fig:IC_Mag_a3}(c) and \ref{fig:IC_Mag_a3}(d). Fig.~\ref{fig:IC_Mag_a3}(c) shows scans for the peak ($-q_{\rm h}$ $1-q_{\rm k}$ 0), another peak in the Q1 domain that is the partner magnetic satellite of the ($q_{\rm h}$ $1+q_{\rm k}$ 0) peak [Fig.~\ref{fig:IC_Mag_a3}(b)] about the (010) position. The $P$ dependence of this peak is qualitatively similar to that shown in Fig.~\ref{fig:IC_Mag_a3}(b), with the peak becoming significantly suppressed for $P>$~2.7~GPa. In contrast, Fig.~\ref{fig:IC_Mag_a3}(d) show scans for the ($q_{\rm h}$ $1-q_{\rm k}$ 0) peak, a peak in the Q2 domain that is the partner satellite of ($-q_{\rm h}$ $1+q_{\rm k}$ 0) [Fig.~\ref{fig:IC_Mag_a3}(a)]. The data show the intensity of the ($q_{\rm h}$ $1-q_{\rm k}$ 0) peak to not be suppressed by high $P$, but instead to become larger. This increase could arise, for example, due to a change in the precise magnetic structure in the Q2 IC domain, as already suggested by the monotonic $P$ dependence of the incommensurability. However, the leading $P$ effect on the integrated intensities is more readily attributed to a clear suppression of the Q1 $k$-domain relative to the Q2 $k$-domain for $P>$~2.7~GPa.

\subsection{Magnetic Domains}
\label{sec:3_4_MagDom}

Next we use the integrated intensities of the magnetic peaks to estimate quantitatively how the C and IC magnetic domain populations evolve with $P$. Doing this requires models for the magnetic structures. Since the limited quantity of data collected do not allow for full magnetic structure refinements at each $P$, we are restricted to using the magnetic structure models determined accurately at ambient $P$~\citep{Whi12}. Making this choice introduces some assumptions in our analysis; firstly we neglect any $P$ driven change in the precise magnetic structures which, despite not dominating our observations, is nonetheless evident in the data. Secondly, we can obtain no insight concerning the $P$ evolution of the size of the ordered moment. These factors can act in concert with a $P$ dependence of the domain populations, making us unable to obtain a complete quantitative description of our data. However, since there is no change of the intrinsic symmetries of the magnetic structures up to 5.1~GPa, the aforementioned limitations can be negated by working with \emph{ratios} of integrated intensities for the relevant magnetic peaks. From this approach we can obtain quantitatively reliable estimates for the $P$ evolution of the C and IC magnetic domain populations.

We start first with the C domains. As explained in Ref.~\onlinecite{Whi12}, the mode amplitudes of the active $\Gamma^{1}$ and $\Gamma^{3}$ irreps can be either added or subtracted. This leads to four distinct domains $(++)$, $(--)$, $(+-)$ and $(-+)$, where the first (second) symbol denotes the sign of the modes due to $\Gamma^{1}$ ($\Gamma^{3}$). These four domains can be further divided into two groups. The first group C1 includes the $(++)$ and $(--)$ domains. These amount to a S-domain and its time-reversal counterpart which cannot be distinguished in our experiment~\citep{Bro93}. The second group C2 is composed of the $(+-)$ and $(-+)$ domains which likewise can not be distinguished. Consequently, the structures denoted C1 in Fig.~\ref{fig:Mag_Structs}(d) [the $(++)$ domain], and C2 in Fig.~\ref{fig:Mag_Structs}(e) [the $(+-)$ domain], provide representative models for each of the two groups.

Since the scattering patterns due to the C1 and C2 domain groups overlap in reciprocal space, the total intensity of each magnetic peak will depend on the relative population fraction of each C domain group. According to the reported model for the C order in the MF phase~\citep{Whi12}, for certain peaks where $h\neq0$ and $k\neq0$, such as the $\{110\}$ peaks, the C1 and C2 domain groups scatter with different weights. For example, while the expected ratio of integrated intensities for the $(110)$ and $(−110)$ peaks,$I(110)/I(-110)=$~1 when C1 and C2 domains equally populate the sample, this ratio equals 16.5 if only C2 domains exist in the sample. To obtain the C1 : C2 domain population ratio at each $P$, we modelled the overall scattering at both the (110) and ($-$110) positions due to \emph{both} of the C1 and C2 domains, and adjusted their relative population so that the calculated ratio $I(110)/I(-110)$ agrees with the experimental one deduced from the data shown in Fig.~\ref{fig:C_Mag}. The results of this analysis are shown in Table~\ref{tab:domains}.


For the IC spiral order there are also two domain groups due to the two configurational propagation vectors, $\textrm{Q}_{\rm ic1}$ and $\textrm{Q}_{\rm ic2}$. Representative spiral structures for each group are respectively shown in Figs.~\ref{fig:Mag_Structs}(f) (denoted Q1) and (g) (denoted Q2). Within each $k$-domain there also exist two spiral-handedness domains~\citep{Bro93}. Since these cannot be distinguished in our unpolarised neutron scattering experiment, it suffices to use the structures shown in Figs.~\ref{fig:Mag_Structs}(f) and (g) as models for all possible IC structures in the Q1 and Q2 $k$-domains. Using the integrated intensities for the strongest IC peak from the Q1 domain ($q_{\rm h}$~$1+q_{\rm k}$~0), and the strongest IC peak from the Q2 domain ($-q_{\rm h}$~$1+q_{\rm k}$~0), the relative domain populations are determined by Q1=$I(q_{\rm h}~1+q_{\rm k}~0)/(I(-q_{\rm h}~1+q_{\rm k}~0)+I(q_{\rm h}~1+q_{\rm k}~0))$ and Q2=1$-$Q1~\citep{Not01}. As can be deduced from Table~\ref{tab:domains}, the same results for the domain populations can obtained from the ratio of integrated intensities $I(-q_{\rm h}~1+q_{\rm k}~0)/I(q_{\rm h}~1+q_{\rm k}~0)$.

\begin{table}
\caption{\label{tab:domains} The $P$ dependence of the populations of the C1 and C2 C domain groups, and the Q1 and Q2 IC domain groups using the magnetic structure models determined at ambient $P$ reported in Ref.~\onlinecite{Whi12}. Data for the C domains at 0.8~GPa is not available since the (110) peak was not measured.}
\begin{ruledtabular}
\begin{tabular}{c|cccc}
$\begin{array}{c}{\rm Pressure}\\ {\rm (GPa)}\end{array}$ & $\frac{I(110)}{I(-110)}$ & C1 : C2 & $\frac{I(-q_{\rm h} 1+q_{\rm k} 0)}{I(q_{\rm h} 1+q_{\rm k} 0)}$ & Q1 : Q2 \\
\hline\\
0.8 & - & - & 1.18(3) & 0.46(2) : 0.54\\
2.7 & 0.95(4) & 0.52(2) : 0.48 & 1.03(2) & 0.49(2) : 0.51\\
4.1 & 1.32(5) & 0.40(2) : 0.60 & 1.84(5) & 0.35(2) : 0.65\\
5.1 & 1.77(5) & 0.32(2) : 0.68 & 3.29(10) & 0.23(2) : 0.77
 \\
\end{tabular}
\end{ruledtabular}
\end{table}

\begin{figure}
\includegraphics[width=0.48\textwidth]{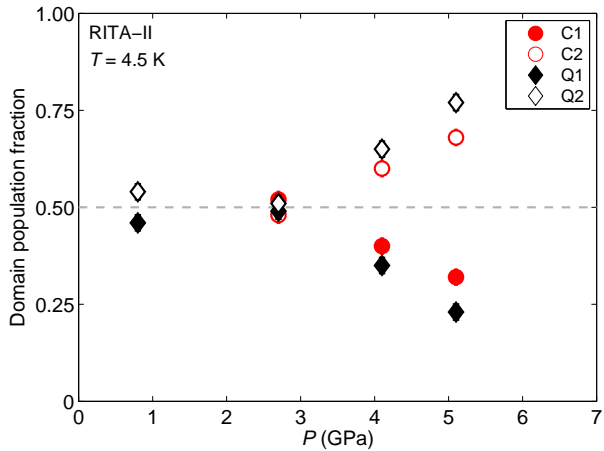}
\caption{(Color online) $P$ dependence of the domain population fraction for the various C and IC domain groups listed in Table~\ref{tab:domains}. The dashed line represents a population fraction of 50~\%, which would be expected for all domain groups in a zero-field-cooled sample at ambient $P$.}
\label{fig:P_dep_domain_pops}
\end{figure}

As reported in Table~\ref{tab:domains}, and plotted in Fig.~\ref{fig:P_dep_domain_pops}, we see that for both C and IC magnetic orders, approximately equal domain populations exist at low $P$. This is as would be expected for the MF phase prepared in a sample at ambient $P$ and zero applied fields. In contrast, with increasing $P>$~2.7~GPa the domain populations for both types of order become unequal. The C order becomes described overall by a larger fraction of domain C2 than C1, and at the same time the IC order by a larger fraction of domain Q2 than Q1.

\section{DISCUSSION}
\label{sec:4Discussion}
The salient results of our high $P$ experiments are as follows:
\begin{enumerate}
\item From room $T$ SXRD experiments a discontinuous crystal structure transition is observed to onset at $P^{\ast}_{\rm RT}=$~7.35(10)~GPa, and take place between the low $P$ $\alpha$-MGO ($Pnma$) phase, and a high $P$ phase identified as a monoclinic ($I2/m$) derivative of the $\beta$-MGO ($Imma$) phase.
\item From the low $T$ neutron diffraction at 4.5~K, measurements up to 5.1~GPa evidence no $P$ induced change in either the lattice or magnetic symmetries compared with at ambient $P$. This is consistent with the survival also up to 5.1~GPa of both bulk FM and FE in the MF phase.
\item The neutron measurements thus confirm the robustness of the multi-$k$ magnetic order in the MF phase, in particular its IC component, up to the highest explored $P$.
\item The neutron study also shows that the near-equal magnetic domain populations for both C and IC modulations observed at lower $P$ is lost as $P>$~2.7~GPa, and the stability of certain C and IC domains becomes enhanced at the expense of others.
\end{enumerate}


We start by discussing points one to three on the above list. Our data show that applying high $P$ leads to a modest, yet clear change in the precise incommensurability of the IC magnetic order [Figs.~\ref{fig:IC_Mag}(e) and (f)]. The larger $P$ dependence is observed for the $q_{\rm h}$ component compared with the $q_{\rm k}$ component, as might be naively expected for the larger lattice compressibility along the \textbf{a}-axis than the \textbf{b}-axis. As mentioned already in Ref.~\onlinecite{Hon14}, fully understanding such behavior requires complementary high $P$ measurements of the magnetic excitation spectrum to determine the $P$ evolution of the interactions. Such a study may be achievable with the high $P$ PE press setup, and making use of developments in instrumentation for inelastic neutron scattering on small samples~\citep{Fre15}.

What is not inferred from our data is that at $P^{\ast}_{\rm LT}$ where FE disappears, the frustrated interactions that lead to IC spiral formation tend towards being fully resolved. If this were the case, we could expect the IC peak intensities to vary \emph{smoothly} towards becoming entirely suppressed at $P^{\ast}_{\rm LT}$, or that the IC propagation vector transforms into a C one that locks-in at $P=P^{\ast}_{\rm LT}$. Based on our data neither of these scenarios seems likely.

Instead, based on the room $T$ SXRD data, the scenario that emerges is that the IC order collapses discontinuously at $P^{\ast}_{\rm LT}\approx$~6~GPa due to a change in crystal symmetry away from the olivine $\alpha$-MGO phase. This conjecture requires that the room $T$ crystal structure transition we observe at $P^{\ast}_{\rm RT}$ to occur similarly at very low $T$s. This expectation finds support from examining the better-determined $P$-$T$ structural phase diagrams for isostructural minerals that display the $P$ driven $\alpha$-$\beta$ transition~\citep{Fei99}. In the case of Mg$_{2}$SiO$_{4}$~\citep{Fei99}, the $\alpha$-$\beta$ transition $P$ decreases as $T$ also decreases, and at a rate quantitatively similar to the one required to explain the present measurements on MGO ($P^{\ast}_{\rm RT}=$~7.35(10)~GPa at $T=$~300~K and $P^{\ast}_{\rm LT}$$\approx$~6~GPa at $T=$~4.5~K). Thus, a structural transition towards a high $P$, possibly monoclinic, $\beta$-MGO phase provides a feasible explanation for the suppression of the FE generated by magnetism in the $\alpha$-MGO phase. This proposal can be tested directly with low $T$, high $P$ studies of the crystal structure, and augmented by further bulk magnetic and electric measurements for a more complete characterisation of the low $T$ phases for $P>$~6~GPa.



Next we discuss the fourth point on the above list of salient results; the observation from high $P$ neutron diffraction that for each of the C and IC orders, the two possible domain groups become unequally populated for $P>$~2.7~GPa. In the absence of applied fields, the only feasible cause for the domain imbalance is due to the existence of finite non-hydrostatic, or uniaxial stress, $P$ components exerted on the sample by the solid Pb PTM. In the present case we can estimate the size of any uniaxial $P$ component, $P_{\rm uni}$ must lie in the range 0~$<$$P_{\rm uni}$$<$~1.0~GPa~\citep{Footnote}. This includes the typical $P$ range of up to a few kbar achievable in dedicated uniaxial $P$ neutron studies~\citep{Nak03,Nak11,Cha15,Naf16}. In our experiment however, the precise sizes and directions of $P_{\rm uni}$ can not be determined.

Despite this, we nevertheless draw analogy between our observations and the common use of uniaxial stress to distinguish between single-$k$ and multi-$k$ magnetic structures~\citep{Ros84}. Namely, we expect the non-hydrostatic $P$ component to be minor compared with the isotropic one, and that it does not itself strongly distort the magnetic structures. Instead, due to the finite magnetoelastic coupling~\citep{Hon12} we expect the stress $P$ components to mainly influence the thermodynamic stability of multi-$k$ domains that would otherwise nucleate with equal probability under truly hydrostatic $P$ conditions.

Evidence to support this hypothesis is seen in Fig.~\ref{fig:P_dep_domain_pops}. For $P>$~2.7~GPa the domain groups C2 and Q2 each become more populated in the sample relative to the domain groups C1 and Q1. In addition, and bearing in mind the assumptions used for the analysis, there is a reasonable quantitative agreement between the rates of the $P$ dependent increases in both the C2 C and Q2 IC domain fractions for $P>$~2.7~GPa, and the concomitant falls in the fractions of C1 and Q1 domains. Using this observation we propose the stress $P$ effect to enhance the stability of multi-$k$ structures composed of superposed domains from the groups C2 and Q2 at the expense of those created from C1 and Q1 domains. It follows that we can propose the existence of a coupling between magnetic order parameters associated with the C2 (C1) and Q2 (Q1) domains, and at the same time the absence of a coupling between the C1 (C2) and Q2 (Q1) domains.


\begin{figure}
\includegraphics[width=0.48\textwidth]{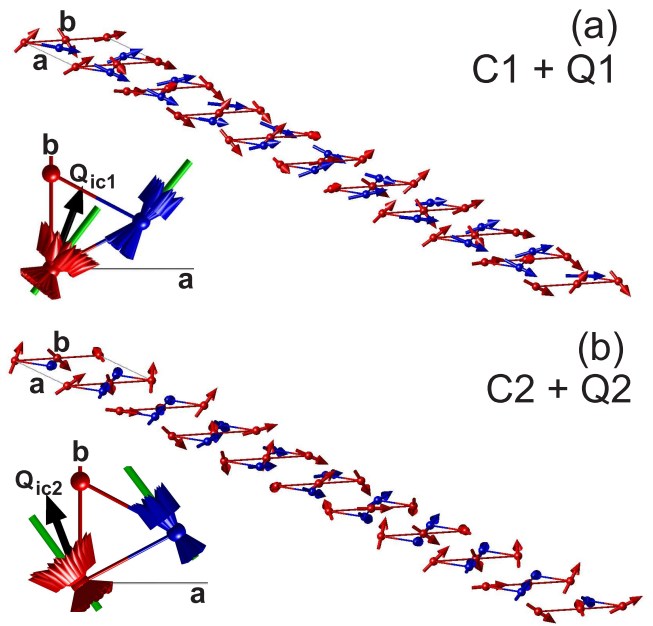}
\caption{(Color online) Deduced multi-$k$ magnetic structures in single MF domains of MGO. The structures are constructed by superposing the representative C and IC structures at ambient $P$ shown in Fig.~\ref{fig:Mag_Structs}. Panel (a) shows the superposition of the magnetic structures for the representative domains from groups C1 and Q1. Panel (b) shows the superposition of the magnetic structures for the domains from groups C2 and Q2. A section of the magnetic structure is shown for a single layer in the \textbf{a}-\textbf{b} plane, and for ten unit cells along the \textbf{a}-axis and one unit cell along the \textbf{b}-axis. The insets to each panel show the calculated magnetic moments (red and blue arrows) across many neighboring sites superposed onto a single site, and viewed along the \textbf{c}-axis. The axes of the resulting cones are shown by green lines. The black arrow denotes the propagation direction of the IC modulation.}
\label{fig:Multik}
\end{figure}

Using the above deductions we next propose complete multi-$k$ magnetic structures that may be realised as the groundstate for the MF phase at ambient $P$. Figures~\ref{fig:Multik}(a) and \ref{fig:Multik}(b) show possible double-$k$ domains constructed after superposing the magnetic structures for the representative domains from groups C1 [Fig.~\ref{fig:Mag_Structs}(d)] and Q1 [Fig.~\ref{fig:Mag_Structs}(f)], and C2 [Fig.~\ref{fig:Mag_Structs}(e)] and Q2 [Fig.~\ref{fig:Mag_Structs}(g)]~\cite{Not02}. Each panel of Fig.~\ref{fig:Multik} shows a section of the overall spin structure in the \textbf{a}-\textbf{b} plane. For each superposition, and each Mn$^{2+}$ site, adjacent spins along the \textbf{a} axis form right conical structures with the cone axes lying almost perfectly within the \textbf{a}-\textbf{b} plane. This is made clearer by the insets for each of Figs.~\ref{fig:Multik}(a) and \ref{fig:Multik}(b). For the superposition C1+Q1 shown in Fig.~\ref{fig:Multik}(a), the cone axes formed by the moments on both Mn$^{2+}$ sites lie at $\approx$+37$^{\circ}$ from the \textbf{b} axis. This lies close to, but not exactly along the direction of the IC modulation vector at ambient $P$, this lying at $\approx$+21$^{\circ}$ from the +\textbf{b} axis. For the superposition C2$+$Q2 shown in Fig.~\ref{fig:Multik}(b), the cone axis lies at $\approx-37^{\circ}$ from the \textbf{b} axis.

Further evidence in support of the above proposed double-$k$ conical structure is obtained from a physical limitation; for each superposition shown in Fig.~\ref{fig:Multik} the net moments on each site are similar in magnitude and always $\leq5\mu_{\rm B}$, as expected for the free ion moment of Mn$^{2+}$ ($S$=5/2). In contrast this limitation becomes violated when attempting other superpositions such as combining domains from the C1 group with the Q2 group. For completeness, triple-$k$ structures involving contributions from multiple C and IC domains were also examined as more complex descriptions of the ambient $P$ MF groundstate. We found that physically allowable triple-$k$ superpositions can be achieved, though the noncollinear Mn moments must then become significantly modulated. In addition, for a triple-$k$ model it becomes challenging to interpret consistently the observed stress-induced tuning of the C and IC peak intensities. For these reasons, we continue our discussion using the above-proposed double-$k$ conical structure as both the simplest and most likely description of the MF groundstate.


\begin{table}
\caption{\label{tab:multik_domains} Table summarizing the eight possible multi-$k$ domains in MGO at ambient $P$, and classified according to their antiferromagnetic degrees of freedom. The FM degree of freedom carried by the C modulation is not considered here explicitly. Thus there are four possible C domains $(++)$, $(--)$, $(+-)$ and $(-+)$, where the first (second) symbol denotes the sign of the modes due to $\Gamma^{1}$ ($\Gamma^{3}$). Each C domain combines only with certain IC domains defined by the IC propagation vector noted in column $\textrm{Q}_{\rm icn}$. In column $h$ the symbols denote the two possible rotation senses of the IC spiral modulation.}
\begin{ruledtabular}
\begin{tabular}{c|cc|cc}
Domain & \multicolumn{2}{c}{C Modulation} & \multicolumn{2}{c}{IC Modulation} \\
No. & $\Gamma^{1}$ & $\Gamma^{3}$ & $\begin{array}{c}\textrm{Q}_{\rm icn}\\ {(n)}\end{array}$ & $h$ \\
\hline\\
1 & $+$ & $+$ & 1 & $+$ \\
2 & $+$ & $+$ & 1 & $-$ \\
3 & $-$ & $-$ & 1 & $+$ \\
4 & $-$ & $-$ & 1 & $-$ \\
5 & $+$ & $-$ & 2 & $+$ \\
6 & $+$ & $-$ & 2 & $-$ \\
7 & $-$ & $+$ & 2 & $+$ \\
8 & $-$ & $+$ & 2 & $-$
 \\
\end{tabular}
\end{ruledtabular}
\end{table}

In Table~\ref{tab:multik_domains} we categorize the distinct types of double-$k$ conical MF domains that may exist at ambient $P$ according to their antiferromagnetic degrees of freedom. The two groups of C domains are included, and denoted as $(++)$ and $(--)$ (group C1), and $(+-)$ and $(-+)$ (group C2) according to the signs of the mode amplitudes due to irreps $\Gamma^{1}$ (first symbol) and $\Gamma^{3}$ (second symbol). Each C domain superposes with an IC modulation to form a multi-$k$ conical structure and, in accordance with our results, only superpositions involving domain groups C1+Q1 and C2+Q2 are realized. In addition, there is a degree of freedom associated with the rotation sense of the IC spiral within each of the Q1 and Q2 domain groups, as denoted by the sign of $h$ in Table~\ref{tab:multik_domains}~\citep{Bro93}. Therefore, in the absence of external perturbation eight domains with distinct antiferromagnetic properties are expected to nucleate with equal probability.

The deduction by empirical means that only certain types of double-$k$ conical domains can be realised in MGO is expected to be symmetry-enforced, and consistent with the invariant terms of the free energy expansion that describes the MF phase~\citep{Har16}. Such a phenomenology will also describe the allowed couplings between applied electric and/or magnetic fields and i) the various types of double-$k$ conical order listed in Table~\ref{tab:multik_domains}, and ii) the bulk FM and FE orders. Indeed, the applied field control of the bulk FM and FE order parameters reported in Ref.~\onlinecite{Whi12} must be reflected by a concomitant control of the underlying conical domain populations. Further neutron studies on single crystal samples can characterise the response of the conical domain populations to applied fields, and ultimately provide definitive insight concerning the coupling between ferromagnetism and ferroelectricity~\citep{Hon16}.

Finally we discuss a possible use for the observed uniaxial stress effect on the multi-$k$ domain populations. From the viewpoint of the bulk ferromagnetic and ferroelectric properties, a MF monodomain state can be created under applied magnetic and electric fields along \textbf{c}~\citep{Whi12}. However, the magnetic order in such a sample may always be divided into a minimum of two distinct parts, since applied fields along \textbf{c} do not restrict the formation of multi-$k$ domains with different IC propagation vectors $\textrm{Q}_{\rm ic1}$ and $\textrm{Q}_{\rm ic2}$. The results of our study indicate that this situation can be further simplified through a combination of both applied magnetic and electric fields \emph{and} a uniaxial stress. For judicious choices of all these experimental parameters, it may be possible to prepare a pure MF, ferromagnetic and ferroelectric monodomain state in a sample, with these orders arising from just a single double-$k$ conical domain.


\section{SUMMARY}
\label{sec:5Summ}
In summary, high pressure ($P$) synchrotron x-ray diffraction (SXRD) and single crystal neutron diffraction experiments have been carried out to determine why the ferroelectricity (FE) observed in the olivine Mn$_{2}$GeO$_{4}$ (MGO) becomes suppressed by pressure ($P$) for $P$s above $P^{\ast}_{\rm LT}$$\approx6$~GPa. From single crystal neutron diffraction measurements for $P$s up to 5.1~GPa, the magnetic order is observed to always remains multi-component, i.e. multi-$k$, with each multi-$k$ domain displaying both a commensurate (C) component that generates ferromagnetism, and an incommensurate (IC) spin spiral component that generates FE. The results show that the general symmetry of the magnetic order underlying the novel bulk multiferroic properties likely remains unchanged all the way up to $P^{\ast}_{\rm LT}$. In combination with the high $P$ SXRD data obtained at room $T$, we argue that the IC order that generates FE collapses discontinuously at $P^{\ast}_{\rm LT}$ due to the occurrence of a structural transition between the low $P$ $\alpha$-MGO (olivine) phase and a high $P$ form of the $\beta$-MGO structural phase that hosts a still unknown form of magnetic order.

At the highest $P$s explored in our neutron diffraction experiments, our data also evidence the emergence of a non-hydrostatic, or uniaxial stress, component of the applied $P$. The uniaxial stress component is observed to tune the stability of different multi-$k$ domains, and from our observations we can infer the existence of a coupling between the C1(C2) and Q1(Q2) order parameters, and absence of a coupling between the C1(C2) and Q2(Q1) order parameters. Based on our observations we propose double-$k$ conical magnetic structures for the multiferroic groundstate, this being a key starting point for any model description of how the multi-$k$ magnetism mediates the coupling between the bulk multiferroic properties. In addition, our observations lead to the expectation that a single type of multi-$k$ domain, a true MF monodomain, can be realized by combining the observed uniaxial stress effect with both applied magnetic and electric fields.

\begin{center}
\textbf{ACKNOWLEDGEMENTS}
\end{center}
Discussions with A.B.~Harris and S.~Klotz are gratefully acknowledged. Neutron experiments were performed at the Swiss spallation neutron source, SINQ, PSI, Switzerland. We are also grateful to the ESRF, Grenoble, France for the allocation of synchrotron beamtime at the SNBL beamline. T.H. acknowledges support from the Condensed Matter Research Center (CMRC) at KEK, Japan.

\bibliography{MGO}

\begin{thebibliography}{43}
\bibitem [{Foo()}]{Footnote}%
  \BibitemOpen
  \href@noop {} {}\bibinfo {note} {As described in Ref.~\onlinecite{Klo12},
  Tresca's plasticity criterion determines the maximum size $P_{\rm uni}$ can
  be as generated by a solid PTM. According to this criterion, the largest
  $P_{\rm uni}$ can be is two times the shear strength, $\tau$ at both the
  relevant applied $P$, and temperature at which the $P$ was changed (since it
  is at this $T$ that any $P$ gradients are induced, so 180~K in our neutron
  experiments). Using data presented in both Refs.~\onlinecite{Klo12} and
  \onlinecite{Tow67}, the $\tau$ for Pb is estimated to be $\sim$0.5~GPa at
  180~K and 5~GPa, thus we estimate $\sim$1.0~GPa as the maximum size $P_{\rm
  uni}$ can take in our experiments. In this context however, it also behoves
  us to consider recent dedicated work wherein shear-induced stresses in solid
  Pb were quantified.~\cite{Wan16} According to these results, $P_{\rm uni}$
  can be expected to be well below the 1.0~GPa limit imposed by the shear
  strength, underlining the good performance of Pb as a solid PTM. Nonetheless,
  it is clear from the present neutron study that a finite $P_{\rm uni}$ indeed
  exists in our setup at high $P$s, leading us to conclude that in our
  experiments $P_{\rm uni}$ always lies within the range 0~$<$$P_{\rm
  uni}$$<$~1.0~GPa.}\BibitemShut {Stop}%
\end{thebibliography}

\end{document}